\documentstyle[12pt]{article}

\begin{document}

\setcounter{page}{100}

\centerline{\large {\bf Multi-color QCD at High Energies and
Exactly Solvable Lattice Theories}}
\vskip.3in
\centerline{ L. N. Lipatov$^{*}$ and A. Berera$^{**}$}
\centerline{ $^{*}$Petersburg Nuclear Physics Institute}
\centerline{ Gatchina, 188350}
\centerline{ St. Petersburg, Russia}
\vskip.3in
\centerline{ $^{**}$Department of Physics}
\centerline{ Pennsylvania State University}
\centerline{ University Park, Pennsylvania 16802, USA}

\vskip.6 in
\centerline{\bf Abstract}
\vskip.1in
We examine the generalized leading-logarithmic approximation
(LLA) equations for compound states of n-reggeized gluons.  It
is shown that in multi-color QCD, when $N_c \rightarrow \infty$,
these equations have a sufficient number of conservation laws
to be exactly solvable.  Holomorphic factorization
of the wave functions is used to reduce the corresponding
quantum mechanical problem to the solution of the one-dimensional Heisenberg
model with the spins being the generators of the M$\ddot{o}$bius
group of conformal transformations.
\vskip.1in

\vskip.3in
{\bf 1. Introduction}
\vskip.1in

This talk is centered around obtaining the exact solution to a perturbative
QCD evolution equation known as the Bartels-Kwiecinski-Praszalowicz
(BKP)-equation$^1$ in the limiting case where the number of colors
of gluons, $ N_c$, is infinite.  One may wonder what relevance any
equations of perturbative QCD may have in understanding the low energy
confining properties of the theory.  The answer to this question
is not well defined.  However, what is clear is that only in the
perturbative regime of QCD, we are able to exactly treat gauge and
Lorentz invariance.  Even then, within this regime one discovers that
such a task is nontrivial.  Thus the first lesson one gains from
examination of perturbative QCD is experience with nonabelian gauge
calculations that can be tested for their correctness.

That may be a
useful reason for those working in
low energy QCD to nevertheless study the high energy
regime as a warm-up exercise.  However that is not the primary reason
for this talk.  The general class of equations that we are considering
here are the only known evolution equations in QCD
that exactly respect gauge invariance and have a kinematic regime in which
they are exactly valid.  Although their practical uses are for
calculating high energy scattering amplitudes, it is
natural to also examine what properties of these equations and their
solutions correspond to what we believe to be true at low energy.

To start with, let us first introduce the names of the equations
to which we are
referring and give some history on their development.  The first
evolution equation in this class was the Gribov-Lipatov-Altarelli-Parisi
(GLAP) equation$^2$ which was derived in the early 70's.
This equation is a predecessor to the main equations we want to discuss here.
 From this group, the first was
the Fadin-Kuraev-Lipatov (FKL)-equation$^3$, which was derived in 1975.
This equation was the initial form of a more contemporary
version known as the Balitsky-Fadin-Kuraev-Lipatov
(BFKL)-equation$^3$.
The FKL-equation
was derived for massive Yang Mills theory with a massive Higgs particle
and for arbitrary SU(N) gauge group.
This is an equation for the two-to-two scattering amplitude
in the Regge limit,
$m^2 e^{1/g^2} \sim s  \gg m^2 \sim t$,
where m is the mass of the vector boson,
$\sqrt{s}$ is the center of mass energy and
$\sqrt{-t}$
is the momentum
transfer.
We note that the Regge limit also implies the leading-log-approximation,
where $g^2 \ln (s/t) \sim 1$ and $g^2 \ll 1$.
To obtain the equation, working
in momentum space using s-channel unitarity
along with analyticity, the amplitude was computed to eighth-order.
{}From this the general form could be deduced into what became the
FKL-equation.
Only the t=0 solution was obtained
in [3].  The solution showed that the amplitude violated the
Froissart bound.  However it should be
realized that the region where this violation
occurs is also beyond the region where the FKL-approximations are
valid.

In 1978 it was verified that there are no infrared divergences in
QCD for
scattering of colorless particles at arbitrary
t in the BFKL-equation. In particular this held at t=0, where the
problem is typically most pronounced.  The solution for arbitrary
t was found in 1986$^4$.  A relevant point for
the present discussion is that the calculation was done in transverse
coordinate representation (or impact parameter space).  In
this representation it was recognized that the BFKL-equation had
two-dimensional M$\ddot{o}$bius invariance.

The shortcoming of the BFKL-equation is that it violates
the unitarity bounds.  To correct this, the suggestive approach is
to consider diagrams with an arbitrary number of reggeized
gluons.  The BFKL-equation only accouts for two reggeized gluons.  The
equation with N reggeized gluons was obtained by Bartels and
by  Kwiecinski and Praszalowicz$^1$.
The purpose of this talk is to examine the solutions
of this equation for $N_c \rightarrow \infty$.  What will be
achieved here is a relation of this equation in the above limit,
to exactly solvable models.  The end result is a reduction of the
problem to a one-dimensional lattice model.

Before turning to the quantitative discussion, let us place into
perspective what contact this development makes with the problem
of confinement.  We have believed since the early seventies that
Yang-Mills theory is plausibly the low-energy limit of an
appropriate string theory.  In the high energy limit, one
may then ask if any aspect of QCD's string-like nature manifests.
There is no known reason from general principles to expect this.
Nevertheless, in light of the results we discuss here, we do find
a string-like remnant of QCD in this limit.

Examining the issue a little further, we next recall that
high energy processes in fact have an intrinsic dependence
on the low energy properties of QCD.  We have known since
the 60's, that the dominant exchange in a high energy collision
at fixed t is the pomeron.  From what little we know about the
dynamical make-up of the pomeron, we suspect it is some sort
of collective excitation made of several gluons plus
perhaps quarks.  Thus to study the low energy regime
of QCD, one way
is to focus of particular bound states and try to derive their
properties from QCD.  However another option is to study
the Regge families of hadrons, such as the pomeron, and
try to calculate their parameters from QCD.
We can not offer any reason why the
latter option is better than the former.  However, the one evident
fact is that we have much better experimental data about Reggeons
than about individual hadrons.  Also from the point of view of light-cone
kinematics, the description of Reggeons is more natural than of individual
particles.  If one accepts this line of reasoning to its fullest extent,
one could imagine calculating masses of hadrons using Reggeon
concepts.  At present we do not have sufficient control on the approximations
involved in our evolution equations to justify such calculations. However
one could assume the radius of convergence for our equations is sufficiently
large to make some sort of estimates. We will not discuss this point
further in this talk.

\vskip.3in
{\bf 2. Evolution Equations}
\vskip.1in

The GLAP equation$^2$,
\begin{eqnarray}
{{dn_i(x)} \over {d \xi}} = - \omega_i n_i(x)+ \sum_j \int^1_x
{{dx'} \over x} \omega_{j \rightarrow i}({x \over {x'}}) n_j(x')
\end{eqnarray}
where,
\begin{eqnarray}
\xi={1 \over c} \ln (1+{\alpha \over \pi} {c} \ln {{Q^2} \over {\mu^2}})
\end{eqnarray}
and
\begin{eqnarray}
\omega_i=\sum_{k} \int_0^1 dx \omega_{i \rightarrow k}(x) ,
\end{eqnarray}
determines the $Q^2$-evolution of the parton distributions $n_i(x)$,
where $x={k_+ \over {p_+}}$ is the ratio of the parton
to hadron longitudinal momentum in the light-cone frame.
The splitting kernels, $\omega_{i \rightarrow k}({x \over {x'}})$, describe
the inclusive probabilities of the parton decay into the opening
phase space $d\xi$.  Mellin transforming in ln${x \over {x'}}$
gives the anomalous dimension matrix $\gamma(j)$ for
the twist two operators in QCD.  For example in the case of pure
gluondynamics,
\begin{eqnarray}
n_g(x)= \int^{\sigma+i\infty}_{\sigma-i\infty}
{{dj} \over {2\pi i}} ({1 \over x})^j e^{\xi \gamma(j)}
\end{eqnarray}
where
\begin{eqnarray}
\gamma(j)= {2 \over {j(j-1)(j+1)(j+2)}} - {1 \over {12}} -
\psi(j-1)+\psi(1)
\end{eqnarray}
and $\psi(j) = {{\Gamma^\prime(j)}\over {\Gamma(j)}}$.
Note that $\gamma(j)=\gamma(1-j)- {\pi} ctg(\pi j)$ and
$\gamma(2)=0$ due to the conservation of the stress tensor $T_{\mu \nu}$.
{}From eq. (4) we obtain that for $x \rightarrow 0$, $n_g(x) \sim
{1 \over x} exp(c \sqrt{\xi \ln {1 \over x}})$.
This implies that total cross-section $\sigma_t(x, Q^2)$, for $\gamma^*
p$ scattering at large energies $\sqrt{s}={Q\over \sqrt{x}}$ grows more
rapidly than any power of lns. This is a violation of the Froissart
bound $\sigma_t < c \ln^2 s$.

At small x, for parton distributions $n_g(x, k_\perp)$ depending
on transverse parton momentum $k_\perp$, one should use the BFKL
equation$^3$,
\begin{eqnarray}
{{dn(x,k_\perp)} \over {dln {1 \over x}}} = 2 \omega (-k^2_\perp)
n(x,k_\perp)+\int d^2k'_\perp K(k_\perp, k'_\perp)n(x,k'_\perp)
\end{eqnarray}
where
$n(x)=\int d^2k_\perp n(x,k_\perp)$ and
\begin{eqnarray}
\omega(-k_\perp^2)= - {g^2 \over {16\pi^3}} N_c \int d^2k'_\perp
{{k^2_\perp} \over {(k-k')^2_\perp k^{\prime 2}_\perp}} .
\end{eqnarray}
where $k^2_{\perp}>0$.
Note that the gluon Regge trajectory $j(-k^2_\perp)$ is related to $\omega$
by $j(-k^2_\perp)=1+\omega(-k^2_\perp)$.
The kernel K for
SU($N_c$) gauge theory is,
\begin{eqnarray}
K(k_\perp, k'_\perp)={g^2 \over {4 \pi}}N_c {1 \over {(k-k')_{\perp}^2}} .
\end{eqnarray}
Observe that the infrared divergences cancel in the right hand
side of eq.(6).

The solution of eq. (6) can be written in the form$^3$,
\begin{eqnarray}
n(x,k_\perp)={1 \over x} \sum^{\infty}_{m=-\infty}
e^{im\phi}\int^\infty_{-\infty}d\nu ({1 \over x})^{\omega(\nu, m)}
k_{\perp}^{2i\nu} c_{m,\nu} ,
\end{eqnarray}
where $c_{m, \nu}$ is determined by the initial conditions for
$n(x,k_\perp)$ at
fixed x, $\phi=arctg({{k_{\perp}^x} \over {k_{\perp}^y}})$, and the
eigenvalue $\omega(\nu, m)$ of the corresponding stationary
 equation is,
\begin{eqnarray}
\omega(\nu,m)= {g^2 \over {2\pi^2}} N_c \int^1_0 {{dy} \over {1-y}}
[y^{{-1+|m|} \over 2} \cos(\nu \ln y)-1]=
{g^2 \over {2\pi^2}} N_c (\psi(1)-Re \psi({1 \over 2} +i\nu+
{{|m|} \over 2})) .
\end{eqnarray}
The biggest value of $\omega$ is $\omega(0,0)={g^2 \over {\pi^2}} N_c \ln 2$,
and therefore from eq. (9) we obtain that $n(x,k_\perp) \sim
{1 \over x} ({1 \over x})^{\omega(0,0)}$.  This means that the solution of
the BFKL-equation also does not agree with the Froissart bound.
For this equation as well as for the GLAP-equation, the reason
for this violation is that the evolution equations were obtained in
the leading logarithmic approximation,
where the S-matrix does not satisfy unitarity$^3$.

Thus we find in both cases, the GLAP and BFKL equations, the result is
incomplete.  As such we will construct a modified leading logarithmic
approximation (LLA) that is compatible with the Froissart bound.

\vskip.3in
{\bf 3. Partonic Wave Functions}
\vskip.1in

The partonic distributions $n_i(x, k_\perp)$ are proportional to the imaginary
part of the scattering amplitude at the momentum transfer q=0.  It is
natural to generalize the evolution equations for arbitrary
momentum transfer.  In this case the resulting equations could be
considered as equations
for the hadronic wave function.   The usual Schr$\ddot{o}$dinger
equation $E \psi = H \psi$ determines the mass of the hadron as a function
of its spin, $m^2=m^2(j)$.  To determine j=j($m^2$), one
can replace this equation by the BFKL equation
\begin{eqnarray}
2 H_{12} \psi=E \psi,
\end{eqnarray}
where
\begin{eqnarray}
E=- 16 {{\omega \pi^2} \over {g^2 N_c}}
\end{eqnarray}
and $j=1 + \omega$ is the position of the j-plane singularity of the
t-channel partial wave.  The high energy asymptotics of scattering
amplitudes are determined by the eigenvalues of equation (11)
as $A(s,t) \sim s^{1+\omega(t)}$.  The eigenvalues $\omega$
could in general
also depend on t=$-q^2$, but due to the conformal invariance
of the BFKL equation$^3$, in LLA this dependence is absent.
The operator $H_{12}$ on the left hand side of eq. (11) is$^4$,
\begin{eqnarray}
H_{12}={1 \over {|P_1|^2|P_2|^2}}P^*_1P_2 \ln |\rho_{12}|^2P_1P^*_2
+ h.c.+ \ln (|P_1|^2|P_2|^2)-4\psi(1) ,
\end{eqnarray}
where $\rho_{12}=\rho_1 - \rho_2$,
$\rho_r=x_r+iy_r$, the
momenta $P_r = i {{\partial} \over {\partial \rho_r}}$, and h.c means
the complex conjugated expression.

To unitarize the results of the LLA, one must generalize eq.(11) for compound
states with an arbitrary number of gluons.  Such a generalization was
done in [1].  Here we discuss the BKP-equation for the large
$N_c$ case.  Thus we consider the equation,
\begin{eqnarray}
H \psi=E \psi
\end{eqnarray}
with E as given in eq. (12) and where the Hamiltonian H contains
only interactions of neighboring particles,
\begin{eqnarray}
H= \sum_{r=1}^m H_{r,r+1} .
\end{eqnarray}
The pair Hamiltonian $H_{r,r+1}$ acts on the coordinates r and r+1
of the gluons as given by eq. (13).

Note that there is also a generalization of the GLAP-equation (1) for matrix
elements of quasipartonic operators of high twist$^5$.  This B'F'KL-equation
is also simplified in the region of large $N_c$. It takes the form
of eqs. (14) and (15) with the pair kernel describing the evolution
of twist two operators.
The eigenvalues of this equation are proportional to the anomalous
dimensions of the quasi-partonic operators whose contributions
are important in the small-x region.
We will not discuss this equation any further below.

\vskip.3in
{\bf 4. The BKP-Equation in the Large $N_c$ Limit}
\vskip.1in

{}From eqs. (13) and (14) one can derive the holomorphic separability
of the Hamiltonian, which is a central property for our present
discussion.  Thus we can write
\begin{eqnarray}
H=H + H^* ,
\end{eqnarray}
where $ H$ and $H^*$ act on the holomorphic ($\rho_j$) and
antiholomorphic ($\rho^*_j$) coordinates respectively
with
\begin{eqnarray}
H=\sum_{j=1}^n H_{j, j+1} ,
\end{eqnarray}
and similarly for $H^*$.  The pair holomorphic Hamiltonian is,
\begin{eqnarray}
H_{j,j+1}={1 \over {P_j}} \ln (\rho_{j,j+1})P_j+
{1 \over {P_{j+1}}} \ln (\rho_{j,j+1})P_{j+1}+ \ln (P_j,P_{j+1})-2\psi(1).
\end{eqnarray}
An important outcome of holomorphic separability is that the solution
of eq. (14) separates as$^6$,
\begin{eqnarray}
\psi(\vec{\rho_1},\vec{\rho_2}, \cdots, \vec{\rho_n})=
\sum \psi(\rho_1,\rho_2,\cdots,\rho_n) \tilde{\psi}(
\rho^*_1,\rho^*_2,\cdots,\rho^*_n)
\end{eqnarray}
where the sum is over all degenerate solutions of the Schr$\ddot{o}$dinger
equation in the holomorphic and antiholomorphic subspaces,
\begin{eqnarray}
E=\epsilon + \tilde{\epsilon}, H \psi=\epsilon \psi,
H^* \psi = \tilde{\epsilon} \psi
\end{eqnarray}
The pair Hamiltonian $H_{j,j+1}$ in eq. (18) can also be written in
the forms$^6$,
\begin{eqnarray}
H_{j,j+1} & = & \ln (\rho^2_{j,j+1}P_j)+ \ln (\rho^2_{j,j+1}P_{j+1})-
2 \ln \rho_{j,j+1}
-2\psi(1)\\
& = & \rho_{j,j+1} \ln (P_jP_{j+1})\rho^{-1}_{j,j+1}+
2 \ln \rho_{j,j+1}
-2\psi(1)
\end{eqnarray}
 From the above representations, it is obvious that H is invariant
under the  M$\ddot{o}$bius transformations$^4$,
\begin{eqnarray}
\rho_j \rightarrow {{a\rho_j + b} \over {c\rho_j + d}},
\end{eqnarray}
where a,b,c, and d are arbitrary complex parameters.
The generators of these transformations are
\begin{eqnarray}
\vec{M}=\sum_{i=1}^n\vec{M}_i, \,\,
M_i^z=\rho_i\partial_i \,\,
M_i^-=\partial_i \,\,
M_i^+=-\rho_i^2 \partial_i
\end{eqnarray}
One can also obtain the transposed Hamiltonian $H^T$ from H by
two different similarity transformations,
\begin{eqnarray}
H^T & = & P_1P_2 \cdots P_n HP_1^{-1}P_2^{-1}\cdots P_n^{-1} \\
& = & \rho_{12}^{-1}\rho_{23}^{-1}\cdots\rho_{n1}^{-1} H
\rho_{12}\rho_{23}\cdots\rho_{n1}.
\end{eqnarray}
This implies that there are two different normalization conditions
for the solutions of eq. (14) which are compatible with eqs. (25)
and (26).
These are,
\begin{eqnarray}
||\psi_1||^2=\int \psi^* \prod _{r=1}^n d\rho_r  P_r \, \psi \\
|| \psi_{2} ||^2 = \int \psi^* \prod_{r=1}^n {{d\rho_i} \over
{\rho_{r,r+1}}} \, \psi.
\end{eqnarray}
{}From eqs. (25) and (26) we conclude that there is a nontrivial differential
operator$^6$,
\begin{eqnarray}
A=\rho_{12}\rho_{23} \cdots \rho_{n1} P_1 P_2 \cdots P_n,
\end{eqnarray}
which commutes with H,
\begin{eqnarray}
[A,H]=0
\end{eqnarray}
Below we will show that there are an infinite number of operators
that commute with H.

\vskip.3in
{\bf 5. Equivalence Between Multi-color QCD at High Energies
and an Exactly Solvable Spin Model}
\vskip.1in

We can write down the operator A in eq.(29) as follows$^7$,
\begin{eqnarray}
A= i^n tr(M_1 M_2 \cdots M_n) ,
\end{eqnarray}
where $M_i$ is the 2*2 matrix constructed from the M$\ddot{o}$bius
generators $M_i$ in eq. (24),
\begin{eqnarray}
M_i=\left( \begin{array}{cc}
\rho_i\partial_i & \partial_i \\
-\rho^2_i\partial_i & -\rho_i\partial_i \end{array} \right)
\end{eqnarray}
In representation (31) the operator A can be interpreted as a transfer
matrix for a lattice theory.  On the links in the "space" direction
(the auxiliary subspace)
there are discrete variables $\xi$ taking values $\xi = \pm 1$ and
on the links in the "time" direction (the quantum subspace), there
are continuous variables $\rho$.

To solve eq (20) exactly, one should find the one parameter family
of integrals of motion, including the operator A of eq.(31). It turns
out$^7$
that such a family is the following,
\begin{eqnarray}
t(\theta)=tr(L_1(\theta) L_2(\theta) \cdots L_n(\theta)) ,
\end{eqnarray}
where,
\begin{eqnarray}
L_i(\theta) = \left(\begin{array}{cc}
\theta + \rho_i\partial_i & \partial_i \\
-\rho^2_i \partial_i & \theta-\rho_i\partial_i
\end{array} \right)
\end{eqnarray}
is the so called L-operator. Let us also introduce the monodromy matrix,
\begin{eqnarray}
T(\theta)=L_1(\theta) L_2(\theta) \cdots L_n(\theta).
\end{eqnarray}
One can verify$^8$ that it satisfies the following Yang-Baxter equation:
\begin{eqnarray}
T^{i_1 i_{1'}}(u) T^{i_2 i_{2'}}(v)(u - v-P_{12})=(u-v+P_{12})
T^{i_2 i_{2'}}(v)T^{i_1 i_{1'}}(u) ,
\end{eqnarray}
where $P_{12}$ is the operator that interchanges the matrix spin
indices (the right and left ones correspondingly).
By taking the traces over indices $i_r$ and $i_{r'}$, we obtain:
\begin{eqnarray}
t(u)t(v)=t(v)t(u),
\end{eqnarray}
so that the operators defined in eq. (33) commute with each other.

Now we want to prove that the operator $t(\theta)$ of eq. (33) also
commute with the holomorphic Hamiltonian in eq. (20).
For this purpose the idea we use is$^9$ that the spin model with the
transfer matrix (33) can be considered as a modification of the Heisenberg
model.  However, instead of the fundamental representation of the group
SU(2) with spin S=$1 \over 2$, here we have
the infinite-dimensional representation of the M$\ddot{o}$ebious
group SU(1,1) with spin S=0 .  For this new spin model, there is an unique
Hamiltonian
describing the interaction of nearest spins for which the model is exactly
solvable.  The general method of obtaining this Hamiltonian was developed
many years ago$^8$.  Briefly, to do this one should construct the
operator $L^{12}(\theta)$, which satisfies the trilinear Yang-Baxter
equation for the case when both the quantum and auxiliary subspaces are
one-dimensional ($\rho_1$ and $\rho_2$).  Then for this new
spin model, the Hamiltonian is given by eq. (17), where $H_{1,2}$ can be
calculated from the small-$\theta$ expansion of $L^{1,2}(\theta)$:
\begin{eqnarray}
L^{1,2}(\theta)=P^{1,2}(1+\theta H_{1,2} + \cdots).
\end{eqnarray}
Here $P^{1,2}$ is the operator which interchanges the coordinates $\rho_1$
and $\rho_2$.  According to the general theory$^8$, $L^{1,2}(\theta)$
should also satisfy the linear Yang-Baxter equation:
\begin{eqnarray}
L_1(u)L_2(v)L^{1,2}(v-u)=L^{1,2}(v-u)L_2(v)L_1(u)
\end{eqnarray}
In this equation, the $L_i$ operators are 2*2 matrices (34).  From eq. (39)
we find$^3$
that $H_{1,2}$ depends only on the Casimir operator
of the conformal group.  This can be written in the form:
\begin{eqnarray}
(\vec{M}_1+\vec{M}_2)^2=\hat{m}(\hat{m}-1)
\end{eqnarray}
We also find that $H_{1,2}$ satisfies the equation
\begin{eqnarray}
[H_{12}(\hat{m})-H_{12}(\hat{m}-1)](\hat{m}-1)=2,
\end{eqnarray}
for which the solution is
\begin{eqnarray}
H_{1,2}=\psi(\hat{m})+\psi(1-\hat{m})-2\psi(1)
\end{eqnarray}
up to an additional term $\Delta(\hat{m})$, which is a periodic
function (ie. $\Delta(\hat{m})=\Delta(\hat{m}+1)$).
Using eq. (10), we can verify that the expression for $H_{1,2}$ determined
by eqs. (18) and (42)
coincide.  Thus, according to the general theory in [8],
the Hamiltonian (17) commutes with all operators of the type
$t(\theta)$ in eq.(33).

\vskip.3in
{\bf 6. Conclusion}
\vskip.1in

We have shown above that in the generalized leading logarithmic approximation,
the equation for the compound states of n-reggeized gluons
is significantly simplified in the large $N_c$-limit.  In particular,
it is conformally invariant and the Hamiltonian has the remarkable
property of holomorphic separability.  In addition, the equations
for holomorphic and antiholomorphic wave functions have a sufficient
number of conservation laws to be exactly solvable.  This is
related with the fact that the Hamiltonians in the corresponding
subspaces coincide with the local Hamiltonians of the exactly solvable
Heisenberg model for spin S=0.  As such, the quantum mechanical problem
posed in eq. (20) is reduced to the pure algebraic one of constructing
the representations of the Yang-Baxter algebra  in eq. (36). The
simple method of finding these representations was developed in [10].
It is based  on the solution of the Baxter equation
\begin{eqnarray}
\Lambda(\lambda) Q(\lambda)=(\lambda+i)^nQ(\lambda+i)+(\lambda-i)^n
Q(\lambda-i),
\end{eqnarray}
where n is the number of reggeized gluons, $\Lambda(\lambda)$
are the eigenvalues of the operator $t(i\lambda)$ in eq. (33)
and the function Q is the integer function in the complex $\lambda$-
plane.  The eigenvalue $\Lambda(\lambda)$ has the polynomial expansion
in $\lambda$,
\begin{eqnarray}
\Lambda(\lambda)= 2\lambda^n - j(j+1) \lambda^{n-2} + \cdots + A,
\end{eqnarray}
where n is the number of reggeized gluons,
m=j-1 is the conformal weight of the corresponding
composite operator, j(j+1) is the eigenvalue of the Casimir operator
$(\sum_i{\vec{M}_i})^2$, and A is the eigenvalue
of the integral of motion A.  The eigenvalues and eigenfunctions of eqs. (20)
can be expressed through  $Q(\lambda)$.  For n=2
eq. (43) is solved in terms of hypergeometric functions.
For n=3 the solution of eq. (43) for integer j
can be expanded as a linear combination of its solutions $Q_j^{(2)}$
for n=2 as,
\begin{eqnarray}
Q(\lambda)=\sum_{k=1}^j d_k(A) Q_k^{(2)} (\lambda),
\end{eqnarray}
where the parameter A is determined in eq.(44) and $d_k(A)$ are orthogonal
polynomials satisfying the recurrence relations,
\begin{eqnarray}
A d_k(A)= {{k(k+1)} \over {2k+1}}(k-j)(k+j+1)(d_{k+1}(A)+d_{k-1}(A)) .
\end{eqnarray}
The quantization condition for the eigenvalues A is,
\begin{eqnarray}
d_j(A)=0.
\end{eqnarray}
It is possible to express the energies $\epsilon$ in eq. (20) directly in
terms of $d_k(A)$ , when eq.(46) is analytically continued to complex j.
The solution of eq. (46)
would give a possibility to find the Odderon intercept$^{11}$.

\vskip.1in
{\bf Acknowledgment}
\vskip.1in

L.L. is grateful to the Alexander von Humboldt Foundation for
the award which gave him a possibility to work on this
problem and for partial support to INTAS and the Russian Fund of the
Fundamental Investigations (grant 93-02-16809).
Partial support to A.B. was provided by the U. S. Department of Energy grant
no. DE-FG03-91ER40674.

\vskip.3in
{\bf References}
\vskip.1in

1.  J. Bartels, Nucl. Phys. {\bf B175}, (1980) 365;

J. Kwiecinski and M. Praszalowicz, Phys. Lett {\bf B94}, (1980) 413.

2. V. N. Gribov and L. N. Lipatov, Sov. J. Nucl. Phys.
{\bf 15}, (1972) 438;

L. N. Lipatov, Sov J. Nucl. Phys.. {\bf 20}, (1975) 93;

G. Altarelli and G. Parisi, Nucl. Phys. {\bf B26}, (1978) 298.

3. V. S. Fadin, E. A. Kuraev, and L. N. Lipatov, Phys. Lett. {\bf B60},
(1975) 50;

I. I. Balitsky and L. N. Lipatov, Sov. J. Nucl. Phys. {\bf 15} (1978) 438.

4. L. N. Lipatov, Sov. Phys. JETP {\bf 63}, (1986) 904.

5. A. Bukhvostov, G. Frolov, E. Kuraev, and L. N. Lipatov,
Nucl. Phys (1995).

6. L. N. Lipatov, Phys. Lett. {\bf B309}, (1993) 394.

7. L. N. Lipatov, Padova preprint, DFPD/93/T14/70, October 1993.

8. V. A. Tarasov, L. A. Takhtajan, and L. D. Faddeev,

Sov. J. Theor. Math Phys. {\bf 57}, (1983) 163.

9. L. N. Lipatov, JETP Letters {\bf 59}, (1994) 571.

10. L. D. Faddeev and G. P. Korchemsky, Stony Brook preprint,

ITP-SB-94-14, April 1994.

11.  P. Gauron, L. Lipatov, and B. Nicolescu, Phys. Lett {\bf B304},
(1993) 334.

\end{document}